\documentclass[aps,pra,twocolumn,showpacs,nofootinbib,longbibliography,notitlepage]{revtex4-2}
\usepackage{etex}
\usepackage{amsmath,amssymb,amsthm}
\usepackage[colorlinks=true,citecolor=blue,urlcolor=blue]{hyperref}
\usepackage[pdftex]{graphicx}
\usepackage{times,txfonts}
\usepackage{braket}
\usepackage{color}
\usepackage{natbib}
\usepackage{amsmath,blkarray}
\usepackage{mathtools}
\usepackage{latexsym}
\usepackage{tabularx, booktabs}
\usepackage{graphics,epstopdf}
\usepackage{graphicx}
\usepackage{float}
\usepackage{graphicx}
\usepackage{amsfonts}
\usepackage{subcaption}
\usepackage{color,soul}

\newcommand{\be}{\begin{equation}}
\newcommand{\ee}{\end{equation}}
\newcommand{\ba}{\begin{eqnarray}}
\newcommand{\ea}{\end{eqnarray}}

\begin{document}
	\title{Generalized $n$-locality inequalities in star-network configuration and their optimal quantum violations}
		\author{Sneha Munshi}
			\author{Rahul Kumar}
	\author{ A. K. Pan }
	\email{akp@nitp.ac.in}
	\affiliation{National Institute of Technology Patna, Ashok Rajpath, Patna, Bihar 800005, India}
	
	\begin{abstract}
	Network Bell experiments reveal a  form of nonlocality conceptually different from  standard Bell nonlocality. Standard multiparty Bell experiments involve a single source shared by a set of observers. In contrast, network Bell experiments feature multiple independent sources, and each of them may distribute physical systems to a set of observers who perform randomly chosen measurements. The $n$-locality scenario in star-network configuration involves $n$ number of edge observers (Alices), a central observer (Bob), and $n$ number of independent sources having no prior correlation. Each Alice shares an independent state with the central observer Bob. Usually, in network Bell experiments, one considers that each party measures only two observables. In this work, we propose a non-trivial generalization of $n$-locality scenario in star-network configuration, where each Alice performs some integer  $m$ number of binary-outcome measurements, and the central party Bob  performs $2^{m-1}$ binary-outcome measurements. We derive a family of generalized $n$-locality inequalities for any arbitrary $m$. Using {blue}{an elegant} sum-of-squares approach, we derive
	the optimal quantum violation of the aforementioned inequalities can be attained when each and every Alice measures $m$ number of mutually anticommuting observables. For $m=2$ and $3$, one obtains the optimal quantum value {blue}{for qubit system local to each Alice, and it is sufficient to consider the sharing of} a two-qubit entangled state between each Alice and Bob. We further demonstrate that the optimal quantum violation of $n$-locality inequality for any arbitrary $m$ can be obtained when every Alice shares $\lfloor m/2\rfloor$ copies of two-qubit maximally entangled state with the central party Bob. We also argue that for $m>3$, a single copy of a two-qubit entangled state may not be enough to exhibit the violation of $n$-locality inequality but multiple copies of it can activate the quantum violation. We discuss the implications of our study and raise some open questions. 
	\end{abstract}
	\maketitle
	\section{Introduction} 
	
	 Bell's theorem \cite{bell} lies at the heart of quantum foundations.{blue}{ This no-go proof asserts that any ontological model satisfying the locality condition cannot account for all quantum  statistics}.  Apart from its immense impact in quantum foundations research, this theorem paved the path for the development of cutting-edge quantum technologies (see review \cite{brunnerreview}).

	In a conventional Bell experiment, a source distributes a physical system to a set of observers who randomly perform measurements on the respective sub-systems	in their possession. 	The simplest bipartite Bell scenario consists of two distant parties,  Alice and Bob, who apply respective measurements $x$ and $y$, producing outcomes $a$ and $b$. In classical physics, the  system is represented by a classical random variable $\lambda$, which predetermines the local outcomes (reality) independent of the other party's settings and outcomes (locality). The joint probability distribution of the outcomes $a$ and $b$ can then be written as 
	\begin{equation}
	\label{jp}
	P(a,b|x,y)=\int \rho(\lambda)P(a|x,\lambda)P(b|y,\lambda)d\lambda
	\end{equation} 
	where $\rho(\lambda)$ is the probability distribution of $\lambda$. In quantum theory, source can produce an entangled quantum state. In such a case if Alice and Bob perform the measurements of locally incompatible observables, not every joint probability in quantum theory admits the factorized form as in Eq. (\ref{jp}). This feature is referred to as quantum nonlocality and is commonly witnessed via the quantum violation of a suitable Bell's inequality \cite{CHSH1969,brunnerreview}.            
	
	The conventional  multipartite nonlocality is a straightforward generalization of bipartite Bell nonlocality. The study of multipartite nonlocality has been  a vibrant  research area  for last two decades \cite{brunnerreview,Horodecki2009,Guhnea}. In standard multipartite Bell experiments,  three or more distant observers share an entangled state distributed by a single source. A novel approach was recently proposed \cite{Bran2010,Cyril2012} through the network Bell experiments which  demonstrate a form of multipartite nonlocality that conceptually goes beyond the conventional multipartite Bell nonlocality.  In contrast to the standard Bell scenario, the network Bell experiments feature many node observers who hold physical systems originating from different sources. Notably, the sources are assumed to be independent of each other, and therefore \emph{a priori} share no correlations among them. There may be different topological structures of the network, and quantum correlation across the network would manifest in various possible ways. 
	
The simplest network scenario is the bilocality scenario  \cite{Bran2010,Cyril2012}  involving three parties and two independent sources, as depicted in Fig.1. Two edge parties, Alice$_{1}$ and Alice$_{2}$, each share an independent physical system with the central party Bob.  In quantum theory, each source can produce an entangled pair of particles. If Bob performs suitable joint measurement on the two particles in his possession, entanglement can be developed between the particles of two distant edge parties Alice$_{1}$ and Alice$_{2}$. Such a protocol is widely known as entanglement swapping  \cite{Zukw1993}. The correlations between Alice$_{1}$ and Alice$_{2}$ can then violate traditional Bell inequalities upon postselecting on the outcome of the joint measurement by the central party Bob. However, there exists nonlinear inequalities that witness a form of quantum nonlocality of the tripartite correlation in a network devoid of the postselection, known as bilocality inequalities \cite{Bran2010,Cyril2012}. Such inequalities are violated in quantum theory, thereby exhibiting quantum non-bilocality.  It has also been shown that such inequalities  are violated  for all pure entangled quantum states \cite{Gisin2017}.
	
The networks beyond the bilocality scenario feature many independent sources, and each of them distributes physical systems to a set of observers who perform randomly chosen measurements. Despite the initial independence of different sources, a suitably chosen set of measurements can give rise to nonlocal quantum correlations across the whole network. In recent times, the nonlocal quantum correlations in networks have been studied for various topological configurations \cite{Tavakoli2016,Tava2016,Frit2016,Rosse2016,Chav2016,Tava2017,Andr2017,Fras2018,Luo2018,Lee2018,Gupta2018,Wolfe2019,Cyril2019,Renou2019,Kerstjens2019,Aberg2020,Banerjee2020,Gisi2020,Supic2020,Tavakoliarxiv,kundu2020,Gisinarxiv2021,Kraftarxiv,scarani,cont}.  In the triangle network, an interesting  form of genuine quantum nonlocality   is demonstrated \cite{Renou2019} without any inputs, only by considering the output statistics of fixed measurement settings.  The network scenario also allows for nonlocality activation \cite{scarani} and less stringent detection efficiencies\cite{Kerstjens2019}.  It has been shown that an arbitrarily small level of independence is capable of revealing the quantum nonlocality in networks \cite{Supic2020}. The potential of exploiting quantum networks for device-independent information processing has also been discussed \cite{Lee2018}.

 One of the well-studied generalizations of the bilocality scenario is the $n$-locality scenario in star-network configuration \cite{Frit2012,Armi2014}, involving $n$ independent sources. Each source distributes a state to one of the $n$ edge parties (Alices) and the central party (Bob). Experimental  tests of  the bilocality inequality \cite{Saunders2017,Andreoli2017,Carvacho2017,Sun2019} and  the $n$-locality inequalities in star-network configuration \cite{Poderini2020} have also  been reported. The $n$-locality in the star-network scenario is commonly studied for two binary-outcome observables (say, $m=2$) per party.

In this work, we provide a non-trivial generalization of the $n$-locality scenario in star-network configuration, as depicted in Fig.2.   Instead of two observables per party, we consider that each of the $n$ numbers of Alices performs an arbitrary $m$ number of binary-outcome measurements, and Bob performs $2^{m-1}$ binary-outcome measurements. We then derive a family of generalized $n$-locality inequalities for arbitrary $m$ and demonstrate optimal quantum violations. We further show that for any given $m\geq 2$, the optimal quantum violation can be attained when every Alice chooses $m$ number of mutually anticommuting observables. 

We note here that, for $m=2$,  the optimal quantum violation of $n$-locality inequality has been achieved when each Alice shares a two-qubit maximally entangled state with the central party \cite{Andr2017}. We first demonstrate that a two-qubit entangled state can also provide optimal violation for the $m=3$ case. Since the number of mutually anticommuting observables for a qubit system is restricted to at most three, one requires a higher dimensional system for $m>3$. We show that the optimal quantum violation of the family of generalized $n$-locality inequalities is attained if $\lfloor\frac{m}{2}\rfloor$ copies of two-qubit maximally entangled states are shared between each Alice and Bob.  We further argue that for $m>3$, a single copy of a two-qubit entangled state may not violate the generalized $n$-locality inequality, but multiple copies of it may activate the quantum violation. 

%The plan of the paper is the following. In Sec. \ref{II},  we encapsulate the bilocality and $n$-locality scenario when each party performs two measurements ($m=2$). In Sec. \ref{III}, we derive the bilocality inequality for $m=3$ and generalize it for $n$-locality scenario in star network configuration. We introduce our generalized bilocality inequality for arbitrary $m$ in Sec. \ref{IV} and extend it to $n$-locality scenario in star-network configuration.  The optimal quantum violation is also derived in Sec. \ref{IV} We summarize and discuss our results in Sec. \ref{V}.
	\section{ $n$-locality scenario in star-network configuration for $m=2$}
	\label{II}

	Lets us first encapsulate the essence of the simplest network scenario, i.e., the bilocality scenario \cite{Cyril2012} for $m=2$. As depicted in Fig. 1, this network involves  three parties and are two independent sources $S_{1}$ and $S_{2}$. While the source $S_{1}$ sends particles to Alice$_{1}$ and Bob, the source $S_{2}$ sends particles to Alice$_{2}$ and Bob.  Alice$_1$ and Alice$_2$ perform  measurements on their respective sub-systems according to the  inputs  $x_{1}, x_{2}\in\{1,2\}$  producing the outputs $a_{1}, a_{2}\in \{0,1\}$ respectively. Bob performs the measurements on the joint sub-systems   according to the  inputs  $ i\in\{1,2\}$ producing outputs $b\in \{0,1\}$. 
	\begin{figure}[h]
			{{\includegraphics[width=0.95\linewidth]{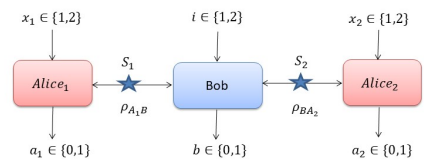}}}
			\caption{Standard bilocality Scenario (See text)  }
			\label{FIG. 1}
		\end{figure}
	Let us now assume that two different hidden variables $\lambda_{1}$ and $\lambda_{2}$ corresponding to the two sources $S_{1}$ and $S_{2}$ respectively, having joint probability distribution $\rho(\lambda_{1},\lambda_{2})$. Since $S_{1}$ and $S_{2}$ are independent, it is natural to assume that the $\lambda_{1}$ and $\lambda_{2}$ are uncorrelated. The joint distribution  $\rho{(\lambda_{1},\lambda_{2})}$ can then be factorized as  $\rho{(\lambda_{1},\lambda_{2})}$ = $\rho_{1}{(\lambda_{1})}$ $\rho_{2}{(\lambda_{2})}$ with  $\int d\lambda_{1}\rho_{1}{(\lambda_{1})}=1$ and $\int d\lambda_{2}\rho_{2}{(\lambda_{2})}=1$. This constitutes the bilocality assumption in a tripartite network scenario \cite{Cyril2012}. Under this assumption, the tripartite joint probability distribution of the outcomes $a_{1}, a_{2}$ and $b$ can be written as
	\begin{eqnarray}
	\label{bl assumption}
	\nonumber
	P(a_{1},b, a_{2}|x_{1},i,x_{2})&=&\int\int d\lambda_{1} d\lambda_{2}\hspace{2mm}\rho_{1}(\lambda_{1})\hspace{1mm}\rho_{2}(\lambda_{2})\\
	&&\times P(a_{1}|x_{1},\lambda_{1})P(b|i,\lambda_{1},\lambda_{2})P(a_{2}|x_{2},\lambda_{2}).\hspace{9pt}
	\end{eqnarray}
	While Alice$_{k}$'s outcomes  solely depend on $\lambda_{k},  k\in\{1,2\}$, note that Bob's outcomes depend on both $\lambda_{1}$ and $\lambda_{2}$. In the standard bilocality scenario ($n=2, m=2 )$, Alice$_{1}$ (Alice$_{2}$) performs two dichotomic measurements $A^{1}_{1}$ and $A^{1}_{2}$ ($A^{2}_{1}$ and $A^{2}_{2}$) corresponding to the input $x_{1}\in\{1,2\}$ ($x_{2}\in\{1,2\}$). Bob  performs the measurement of two observables $B_{1}$ and $B_{2}$ corresponding to the input $i\in\{1,2\}$. It has been proved by Branciard \emph{et al.} \cite{Cyril2012} that the bilocality assumption for $P(a_{1},b,a_{2}|x_{1},i,x_{2})$ holds only if the following nonlinear bilocality inequality 
	\ba
	\label{Delbl22}
	(\Delta_{2}^{2})_{bl}=\sqrt{|{J}^{2}_{2,1}|}+\sqrt{|{J}^{2}_{2,2}|}\leq {2}
	\ea
	is satisfied. Here  $J^{2}_{2,1}$ and $J^{2}_{2,2}$ are the linear combinations of suitably chosen tripartite correlations, defined as 
	\begin{eqnarray}
	\label{bl21}
	J^{2}_{2,1}=\langle(A^{1}_{1}+A^{1}_{2}) B_{1} (A^{2}_{1}+A^{2}_{2})\rangle\\
	\label{bl22}
	J^{2}_{2,2}=\langle(A^{1}_{1}-A^{1}_{2}) B_{2} (A^{2}_{1}-A^{2}_{2})\rangle
	\end{eqnarray}
	where $A^{k}_{x_{k}}$ denotes observable corresponding to the input $x_{k}$ of the $k^{th}$ Alice $(k=1,2)$ and $\langle{A^{1}_{x_1}B_{i}A^{2}_{x_2}}\rangle = \sum_{a_{1},b,a_{2}}(-1)^{a_{1}+b+a_{2}}\hspace{1mm}P(a_{1},b,a_{2}|x_{1},i,x_{2})$. 
		
The assumption of independent sources is crucial to derive the bilocality inequality in Eq.(\ref{Delbl22}). Analogously, in quantum theory, two sources produce entangled independent states. It has been shown \cite{Cyril2012} that the optimal quantum value $(\Delta^{2}_{2})_{Q}^{opt}=2\sqrt{2} >(\Delta_{2}^{2})_{bl}$ can be obtained when Alice$_{1}$'s as well as Alice$_{2}$'s observables are anticommuting. 
		
The $n$-locality scenario in star-network configuration for arbitrary $n$ is one of the straightforward generalizations of bilocality scenario ($n=2$). It involves $n$ independent sources and $n+1$ parties. Each of the $n$ number of edge observers (Alices) shares a physical system with the central observer (Bob), originating from independent sources. Let $A^{k}_{1}$ and $A^{k}_{2}$ be the observables of $k^{th}$ Alice with $k=1,2\ldots n$. The $n$-locality inequality for $m=2$  is given by \cite{Armi2014}
		
		\begin{equation}
		(\Delta_{2}^{n})_{nl}=|{J}^{n}_{2,1}|^{1/n}+|{J}^{n}_{2,2}|^{1/n}\leq {2}
		\end{equation}
		 where
\begin{eqnarray}
J^{n}_{2,1}=\langle(A^{1}_{1}+A^{1}_{2})(A^{2}_{1}+A^{2}_{2})\cdots (A^{n}_{1}+A^{n}_{2})B_{1}\rangle\\
J^{n}_{2,2}=\langle(A^{1}_{1}-A^{1}_{2})(A^{2}_{1}-A^{2}_{2})\cdots (A^{n}_{1}-A^{n}_{2})B_{2}\rangle
\end{eqnarray}

The optimal quantum value $(\Delta_{2}^{n})_{Q}^{opt}=2\sqrt{2}$, i.e., same as bilocality case \cite{Armi2014,Andr2017} which is obtained when every Alice chooses anticommuting observables and shares a maximally two-qubit entangled state with Bob. In this work, we go beyond the $m=2$ case and derive a family of generalized $n$-locality inequalities for arbitrary $m$ and demonstrate their optimal quantum violation.

		\section{ $n$-locality scenario in star-network for $m=3$} 
		\label{III}
		For the sake of better understanding, we first demonstrate the bilocality scenario by considering $m=3$.  Alice$_{1}$ and Alice$_{2}$ now perform three dichotomic measurements, and Bob performs four dichotomic measurements. In this context, we propose a nonlinear bilocality inequality is of the form 
		\begin{equation}
		\label{Delta3}
		(\Delta_{3}^{2})_{bl} = \sum\limits_{i=1}^{4}|{J}^{2}_{3,i}|^{1/2} \leq 6 
		\end{equation} where  $J_{3,i}^{2}$ s (with $i=1,2,3,4$)  are suitaby defined  linear combinations of correlations, can explicitly be written as
		\begin{eqnarray}
		\label{10}
		\nonumber
		&&J_{3,1}^{2}=\hspace{1mm}\langle(A^{1}_{1}+A^{1}_{2}+A^{1}_{3})B_{1}(A^{2}_{1}+A^{2}_{2}+A^{2}_{3})\rangle\\
		\nonumber
		&& J_{3,2}^{2}=\hspace{1mm}\langle(A^{1}_{1}+A^{1}_{2}-A^{1}_{3})B_{2}(A^{2}_{1}+A^{2}_{2}-A^{2}_{3})\rangle\\
		&&J_{3,3}^{2}=\hspace{1mm}\langle(A^{1}_{1}-A^{1}_{2}+A^{1}_{3})B_{3}(A^{2}_{1}-A^{2}_{2}+A^{2}_{3})\rangle\\ 
		\nonumber
		&&J_{3,4}^{2}=\hspace{1mm}\langle(-A^{1}_{1}+A^{1}_{2}+A^{1}_{3})B_{4}(-A^{2}_{1}+A^{2}_{2}+A^{2}_{3})\rangle
		\end{eqnarray}
		Let us first prove the inequality in Eq. (\ref{Delta3}) here.  Using this  bilocality assumption and defining $\langle{A^{1}_{x_{1}}}\rangle_{\lambda_{1}} = \sum_{a_{1}}(-1)^{a_{1}}  P(a_{1}|x_{1},{\lambda_{1}})$, and simirly  $\langle{B_{i}}\rangle_{\lambda_{1},\lambda_{2}}$ and $\langle{A^{2}_{x_{2}}}\rangle_{\lambda_{2}}$, we can write
	\begin{eqnarray}
	\nonumber
	J^{2}_{3,1}=\int\int && d\lambda_{1} d\lambda_{2}\hspace{2mm}\rho_{1}(\lambda_{1})\hspace{1mm}\rho_{2}(\lambda_{2})
	\times(\langle{A^{1}_{1}}\rangle_{\lambda_{1}}+\langle{A^{1}_{2}}\rangle_{\lambda_{1}}+\langle{A^{1}_{3}}\rangle_{\lambda_{1}})\\
	\nonumber 
	&&\langle{B_{1}}\rangle_{\lambda_{1},\lambda_{2}}(\langle{A^{2}_{1}}\rangle_{\lambda_{2}}+\langle{A^{2}_{2}}\rangle+\langle{A^{2}_{3}}\rangle_{\lambda_{2}}).
	\end{eqnarray}
	Since,  $|\langle{B_{1}}\rangle_{\lambda_{1},\lambda_{2}}|\leq{1},$ we can write
	\begin{eqnarray}
	\nonumber
	|J^{2}_{3,1}|&\leq&\int{d}\lambda_{1}\rho_{1}(\lambda_{1})|\langle{A^{1}_{1}}\rangle_{\lambda_{1}}+\langle{A^{1}_{2}}\rangle_{\lambda_{1}}+\langle{A^{1}_{3}}\rangle_{\lambda_{1}}|\\
	\nonumber 
	&&\times\int{d}\lambda_{2}\rho_{2}(\lambda_{2})|\langle{A^{2}_{1}}\rangle_{\lambda_{2}}+\langle{A^{2}_{2}}\rangle_{\lambda_{2}}+\langle{A^{2}_{3}}\rangle_{\lambda_{2}}| \hspace{5mm}
	\end{eqnarray}
	The terms  $|J^{2}_{3,2}|, |J^{2}_{3,3}|$ and $ |J^{2}_{3,4}|$ given in Eq. (\ref{10}) can also be written in similar manner. Using the inequality,  $\sum\limits_{i=1}^{4}\sqrt{r_{i}s_{i}}\leq\sqrt{\sum\limits_{i=1}^{4}r_{i}}\sqrt{\sum\limits_{i=1}^{4}s_{i}}$ (for $r_{i}, s_{i} \geq 0$,$i\in[4]$), we find from  Eq. (\ref{Delta3}) that
\begin{eqnarray}
		\label{delta3fac}
		(\Delta^{2}_{3})_{bl}\leq\sqrt{\int d\lambda_{1}\rho_{1}(\lambda_{1})\delta_{1}} \times\sqrt{\int d\lambda_{2}\rho_{2}(\lambda_{2})\delta_{2}}\hspace{3mm}
		\end{eqnarray}
	
where  $\delta_{1}=\bigg[|\langle{A^{1}_{1}}\rangle_{\lambda_{1}}+\langle{A^{1}_{2}}\rangle_{\lambda_{1}}+\langle{A^{1}_{3}}\rangle_{\lambda_{1}}|+|\langle{A^{1}_{1}}\rangle_{\lambda_{1}}+\langle{A^{1}_{2}}\rangle_{\lambda_{1}}-\langle{A^{1}_{3}}\rangle_{\lambda_{1}}|+|\langle{A^{1}_{1}}\rangle_{\lambda_{1}}-\langle{A^{1}_{2}}\rangle_{\lambda_{1}}+\langle{A^{1}_{3}}\rangle_{\lambda_{1}}|+|-\langle{A^{1}_{1}}\rangle_{\lambda_{1}}+\langle{A^{1}_{2}}\rangle_{\lambda_{1}}+\langle{A^{1}_{3}}\rangle_{\lambda_{1}}|\bigg]$ and , $\delta_{2}=\bigg[|\langle{A^{2}_{1}}\rangle_{\lambda_{2}}+\langle{A^{2}_{2}}\rangle_{\lambda_{2}}+\langle{A^{2}_{3}}\rangle_{\lambda_{2}}|+|\langle{A^{2}_{1}}\rangle_{\lambda_{2}}+\langle{A^{2}_{2}}\rangle_{\lambda_{2}}-\langle{A^{2}_{3}}\rangle_{\lambda_{2}}|+|\langle{A^{2}_{1}}\rangle_{\lambda_{2}}-\langle{A^{2}_{2}}\rangle_{\lambda_{2}}+\langle{A^{2}_{3}}\rangle_{\lambda_{2}}|+|-\langle{A^{2}_{1}}\rangle_{\lambda_{2}}+\langle{A^{2}_{2}}\rangle_{\lambda_{2}}+\langle{A^{2}_{3}}\rangle_{\lambda_{2}}|\bigg]$.

Since all the observables are dichotomic having values $\pm1$, it is simple to check that $\delta_{1}=\delta_{2}\leq 6$. Integrating over $\lambda_{1}$ and $\lambda_{2}$ we obtain $(\Delta^{2}_{3})_{bl}\leq 6$, as claimed in Eq. (\ref{Delta3}).

   The optimal quantum violation of the inequality in Eq. (\ref{Delta3}) is derived as $(\Delta_{3}^{2})_{Q}^{opt}=4\sqrt{3}>(\Delta_{3}^{2})_{bl}$ which requires all three observables of Alice$_1$ (and Alice$_2$) to be mutually anticommuting. The detailed derivation of optimal quantum value using an improved version of the sum-of-square (SOS) approach is placed in Appendix \ref{A}.
		
		The bilocality scenario can be straightforwardly extended to $n$-local scenario. Let $A^{k}_{1}$,$A^{k}_{2},$ and $A^{k}_{3}$ are the observables of $k^{th}$ Alice where $k=1,2 \cdots n$. We can then define
	\ba	
	\nonumber
 &&J^{n}_{3,1}=\langle(A^{1}_{1}+A^{1}_{2}+A^{1}_{3})(A^{2}_{1}+A^{2}_{2}+A^{2}_{3})\cdots (A^{n}_{1}+A^{n}_{2}+A^{n}_{3})B_{1}\rangle,\\
\nonumber
 &&J^{n}_{3,2}=\langle(A^{1}_{1}+A^{1}_{2}-A^{1}_{3})(A^{2}_{1}+A^{2}_{2}-A^{2}_{3})\cdots (A^{n}_{1}+A^{n}_{2}-A^{n}_{3})B_{2}\rangle\\
 &&J^{n}_{3,3}=\langle(A^{1}_{1}-A^{1}_{2}+A^{1}_{3})(A^{2}_{1}-A^{2}_{2}+A^{2}_{3})\cdots (A^{n}_{1}-A^{n}_{2}+A^{n}_{3})B_{3}\rangle,\\
\nonumber
 &&J^{n}_{3,4}=\langle(-A^{1}_{1}+A^{1}_{2}+A^{1}_{3})(-A^{2}_{1}+A^{2}_{2}+A^{2}_{3})\cdots (-A^{n}_{1}+A^{n}_{2}+A^{n}_{3})B_{4}\rangle
\ea
 Using the similar procedure adopted for deriving Eq. (\ref{Delta3}), we find the $n$-locality inequality for $m=3$ is given by
		\begin{equation}
		(\Delta_{3}^{n})_{nl}=\sum\limits_{i=1}^{4}|{J}^{n}_{3,i}|^{1/n}\leq {6}
		\end{equation}
		The optimal quantum violation remain same as $(\Delta_{3}^{2})_{Q}^{opt}$.  We show in the Appendix \ref{A} that the optimal quantum violation of $n$-locality inequality is attained when every Alice chooses mutually anticommuting observables and shares a two-qubit maximally entangled state with Bob.
		
		\section{Generalized bilocality and $n$-locality scenario in star-network for arbitrary $m$}
		\label{IV}
				We first generalize the bilocality scenario for any arbitrary $m$ and derive the bilocality inequality. The two edge parties Alice$_{1}$ and Alice$_{2}$ receive respective inputs  $x_{1}, x_{2}\in\{1,2,3,\cdots,m\}$ producing outputs $a_{1},a_{2}\in\{0,1\}$. Bob receives input $i\in\{1,2,\cdots,2^{m-1}\}$ and produces output $b\in\{0,1\}$. Let us propose a generalized bilocality expression for arbitrary $m$ as
	\begin{equation}
	\label{deltam}
	\Delta^{2}_{m}=\sum \limits _{i=1}^{2^{m-1}}\sqrt{|J^{2}_{m,i}|}
	\end{equation} 
	
	where $	J^{2}_{m,i}$ is the linear combinations of suitable correlations are defined as 
		\begin{equation}
		\label{nbell}
		J^{2}_{m,i}=\bigg(\sum_{x_{1}=1}^{m}(-1)^{y_{x_{1}}^{i}}{A^{1}_{x_{1}}}\bigg) B_{i}\bigg(\sum_{x_{2}=1}^{m}(-1)^{y_{x_{2}}^{i}}{A^{2}_{x_{2}}}\bigg)
	\end{equation}
	Here ${y_{x_{1}}^{i}}$ takes value either $0$ or $1$ and same for ${y_{x_{2}}^{i}}$.  For our purpose, we fix the values of ${y_{x_{1}}^{i}}$ and ${y_{x_{2}}^{i}}$ by invoking the encoding scheme used in Random Access Codes (RACs) \cite{Ambainis,Ghorai2018,AKP2020,Asmita} as a tool. This will fix $1$ or $-1$ values of $(-1)^{y_{x_{1}}^{i}}$ and $(-1)^{y_{x_{2}}^{i}}$ in Eq. (\ref{nbell}) for a given $i$. Let us consider a random variable $y^{\delta}\in \{0,1\}^{m}$ with $\delta\in \{1,2...2^{m}\}$. Each element of the bit string can be written as $y^{\delta}=y^{\delta}_{x_{1}=1} y^{\delta}_{x_{1}=2} y^{\delta}_{x_{1}=3} .... y^{\delta}_{x_{1}=m}$. For  example, if $y^{\delta} = 011...00$ then $y^{{\delta}}_{x_{1}=1} =0$, $y^{{\delta}}_{x_{1}=2} =1$, $y^{{\delta}}_{x_{1}=3} =1$ and so on. Now, we denote  the length $m$ binary strings as $y^{i}$ those have $0$ as the first digit in $y^{\delta}$. Clearly, we have $i\in \{1,2...2^{m-1}\}$ constituting the inputs for Bob. If $i=1$, we get all zero bit in  the string $y^{1}$ leading us $(-1)^{x_{1}^{i}}=1$ for every $x_{1}\in \{1,2 \cdots m\}$. 
	
	An example for $m=2$ could be useful here. In this case, we have $y^{\delta}\in \{00,01,10,11\}$ with $\delta =1,2,3,4$. We then denote $y^{i} \equiv \{y^{1},y^{2}\} \in \{00,01\}$ with $y^{1}=00$ and $y^{2}=01$. This also means  $y^{1}_{x_{1}=1}=0, y^{1}_{x_{1}=2}=0, y^{2}_{x_{1}=1}=0$ and $y^{2}_{x_{1}=2}=1$. Putting those in Eq. (\ref{nbell}), we can recover the expressions $J_{2,1}^{2}$ and $J_{2,2}^{2}$ in Eqs. (\ref{bl21}) and (\ref{bl22}) respectively.
	\begin{figure}[h]
	\hspace*{-0.7cm}
	{{\includegraphics[width=1.2\linewidth]{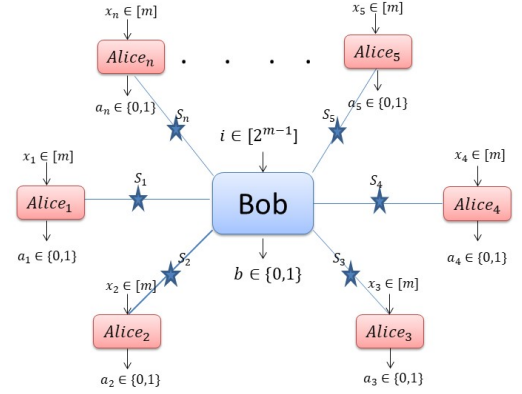}}}
			\caption{$n$-locality scenario in star-network configuration. There are $n$ edge parties (Alice$_k$ with $k\in[n]$) each of them shares a state with the central party Bob. It is assumed that sources $S_{k}$ are independent to each other. Each Alice receives $m$ number of inputs $x_{k}\in [m]$ and Bob receives $2^{m-1}$ number of inputs. }
			\label{FIG. 2}
		\end{figure}

 Following the similar procedure used earlier, by considering the bilocality assumption $\rho(\lambda_{1},\lambda_{2})=\rho_{1}(\lambda_{1})\rho_{2}(\lambda_{2})$,  we can write $J^{2}_{m,i}$ in a hidden variable model as
	\begin{eqnarray}
	\nonumber
	|J^{2}_{m,i}|&=&\int\int d\lambda_{1} d\lambda_{2}\hspace{2mm}\rho_{1}(\lambda_{1})\hspace{1mm}\rho_{2}(\lambda_{2})
	\times\bigg|\bigg(\sum_{x_{1}=1}^{m}(-1)^{y_{x_{1}}^{i}}\langle{A^{1}_{x_{1}}}\rangle_{\lambda_{1}}\bigg)\\
	\nonumber
	&&\hspace{12mm}\bigg(\langle{B_{i}}\rangle_{\lambda_{1}\lambda_{2}}\bigg) \bigg(\sum_{x_{2}=1}^{m}(-1)^{y_{x_{2}}^{i}}\langle{A^{2}_{x_{2}}}\rangle_{\lambda_{2}}\bigg)\bigg|\\
	\nonumber
	&\equiv&\int\int d\lambda_{1} d\lambda_{2}\hspace{2mm}\rho_{1}(\lambda_{1})\hspace{1mm}\rho_{2}(\lambda_{2})
	\times \bigg| J_{m,i}^{2,1}\hspace{1.5mm}\langle{B_{i}}\rangle_{\lambda_{1}\lambda_{2}}\hspace{1mm}J_{m,i}^{2,2}\bigg|
	\end{eqnarray}
	 where to avoid notational clumsiness, we denote
	 \ba
	 \label{Jkmi}
	 J^{2,k}_{m,i}=\sum_{x_{k}=1}^{m}(-1)^{y_{x_{k}}^{i}}\langle{A^{k}_{x_{k}}}\rangle_{\lambda_{1}}
	 \ea
	with $k=1,2$.	Using $|\langle{B_{i}}\rangle_{\lambda_{1}\lambda_{2}}|\leq{1}$ for dichotomic observable and further arranging, we have 
	\begin{eqnarray}
	\label{jmi1}
		\sqrt{|J^{2}_{m,i}|}&\leq& \sqrt{\int d\lambda_{1}\rho_{1}(\lambda_{1})\hspace{1mm}|J^{2,1}_{m,i}|\hspace{0.2mm}\times\hspace{0.2mm}\int d\lambda_{2}\rho_{2}(\lambda_{2})\hspace{1mm}|J^{2,2}_{m,i}|}
	\end{eqnarray}
	 	
	Now, by using Eq. (\ref{jmi1}) and the  inequality $\sum\limits_{i=1}^{2^{m-1}}\sqrt{r_{i}s_{i}}\leq\sqrt{\sum\limits_{i=1}^{2^{m-1}}r_{i}}\sqrt{\sum\limits_{i=1}^{2^{m-1}}s_{i}}$ for $r_{i}, s_{i} \geq 0$, the expression for $\Delta^{2}_{m}$ in Eq.(\ref{deltam}) becomes, 
	
	\begin{eqnarray}
	\label{sumjni}
	\nonumber
	(\Delta^{2}_{m})_{bl}\leq\sqrt{\int d\lambda_{1}\rho_{1}(\lambda_{1})\hspace{1mm}\bigg[\sum_{i=1}^{2^{m-1}}\bigg|J^{2,1}_{m,i}\bigg| \bigg]}
	\times \sqrt{\int d\lambda_{2} \rho_{2}(\lambda_{2})\bigg[\sum_{i=1}^{2^{m-1}}\bigg|J^{2,2}_{m,i}\bigg|\bigg]}\\
	\end{eqnarray}
	
	%\textcolor[rgb]{1,0,0}{Since each $\left\langle A^{1}_{x_{1}}\right\rangle_{\lambda_{1}}$, and $\left\langle A^{2}_{x_{2}}\right\rangle_{\lambda_{2}}\in\{-1,1\}$,} 
	
	We further denote 
	
	\begin{eqnarray}
	\label{alm1}
	\alpha_{m}^{k}=\sum_{i=1}^{2^{m-1}}|J_{m,i}^{2,k}|= \sum_{i=1}^{2^{m-1}}\bigg|\sum\limits_{x_{1}=1}^{m}(-1)^{y_{x_{k}}^{i}}\langle{A^{k}_{x_{k}}}\rangle_{\lambda_{1}}\bigg|
	\end{eqnarray}
	
	where $k=1,2$. Since we have taken the same encoding scheme for Alice$_{1}$ and Alice$_{2}$, then without loss of generality we take $\alpha_{m}^{1}=\alpha_{m}^{2}\equiv\alpha_{m}$. Plugging $\alpha_{m}$ in Eq. (\ref{sumjni})  and integrating over $\lambda_{1}$ and $\lambda_{2}$, we obtain of the bilocality inequality for arbitrary $m$ is 
	\begin{eqnarray}
	\label{Delta2mbl}
	(\Delta_{m}^{2})_{bl}\leq \alpha_{m}.
	\end{eqnarray}
	Hence, the upper bound $\alpha_{m}$ is the key quantity whose maximum value has to be determined for arbitrary $m$. We derive 
	\begin{eqnarray}
	\label{alpham}
		(\Delta_{m}^{2})_{bl}\leq \alpha_{m}=\sum\limits_{j=0}^{\lfloor\frac{m}{2}\rfloor}\binom{m}{j}(m-2j)
		\end{eqnarray}
		The detailed derivation of $\alpha_{m}$ is placed in Appendix \ref{B}. Note that, for $m=2$ and $m=3$ the respective bilocality inequalities $(\Delta_{2}^{2})_{bl}\leq2$ and $(\Delta_{3}^{2})_{bl}\leq 6$ can be recovered. Before demonstrating the quantum violation of bilocality inequality in Eq.(\ref{Delta2mbl}), we derive $n$-locality inequality for arbitrary $m$.
	
In our $n$-locality scenario in star-network configuration we consider that each of the $n$ number of Alices shares a state with Bob, generated by the independent sources $S_{k}$ with $k\in [n]$, as depicted in Fig.\ref {FIG. 2}.  Alice$_k$ performs $m$ number of binary-outcome measurements $A^{k}_{x_{k}}$ ($\forall x_{k}\in[m]$ , for any $k$). Bob receives fixed number of inputs $i \in \{1,2, \cdots,2^{m-1}\}$ and performs binary-outcome measurements on $n$ number of systems he receives from $n$ independent sources. We define the following expression, 
	\begin{eqnarray}
	\label{Deltanmbl}
	(\Delta^{n}_{m})_{nl}=\sum\limits_{i=1}^{2^{m-1}}|J^{n}_{m,i}|^{\frac{1}{n}}
	\end{eqnarray} 
	
	where $J^{n}_{m,i}$ for given $i$, $n$ and $m$ is given by 
	\begin{eqnarray}
	\label{jnn}
	J^{n}_{m,i}= \prod\limits_{k=1}^{n}\bigg[\sum\limits_{x_{k}=1}^{m}(-1)^{y^i_{x_{k}}} A^{k}_{x_{k}}\bigg]\hspace{1pt}B_{i}\equiv \prod\limits_{k=1}^{n} J_{m,i}^{n,k}\hspace{1pt}B_{i}
	\end{eqnarray}
	The term $J_{m,i}^{n,k}$  is same as given in (\ref{Jkmi}).	Using $n$-locality assumption $\rho( \lambda_{1}, \lambda_{2}, \cdots , \lambda_{n})=\prod\limits_{k=1}^{n} \rho_{k}(\lambda_{k})$ and by defining $\langle{A^{k}_{x_{k}}}\rangle_{\lambda_{k}} = \sum_{a_{k}}(-1)^{a_{k}}  P(a_{k}|x_{k},{\lambda_{k}})$, we can write
	\begin{equation}
	\label{Jnmi}
	|J^{n}_{m,i}|^{\frac{1}{n}}\leq \bigg[\prod\limits_{k=1}^{n}\int d\lambda_{k} \rho_{k}(\lambda_{k}) |J^{n,k}_{m,i}|\bigg]^{\frac{1}{n}}
	\end{equation}
	where we consider $|\langle{B_{i}}\rangle_{\lambda_{1},\lambda_{2}..\lambda_{k}}|\leq{1}$. From Eq.(\ref{Deltanmbl}), we get
	\begin{equation}
	\label{deltamn}
	 (\Delta_{m}^{n})_{nl}\leq\sum\limits_{i=1}^{2^{m-1}}\bigg[\prod\limits_{k=1}^{n}\int d\lambda_{k} \rho_{k}(\lambda_{k}) |J^{n,k}_{m,i}|\bigg]^{\frac{1}{n}}
	\end{equation}
	Assuming $\int d\lambda_{k} \rho_{k}(\lambda_{k}) |J^{n,k}_{m,i}|=z_{k}^{i}$ and by using the inequality \cite{Armi2014}
	\begin{equation}
	\label{Tavakoli}
	\ \forall\ \ z_{k}^{i} \geq 0; \ \ \ \sum\limits_{i=1}^{2^{m-1}}\bigg(\prod\limits_{k=1}^{n}z_{k}^{i}\bigg)^{\frac{1}{n}}\leq \prod \limits_{k=1}^{n}\bigg(\sum\limits_{i=1}^{2^{m-1}}z_{k}^{i}\bigg)^{\frac{1}{n}}
	\end{equation} 
	evidently the Eq. (\ref{deltamn}) can be written as
	
	\begin{equation}
	\label{deltamn1}
	 (\Delta_{m}^{n})_{nl}\leq \prod\limits_{k=1}^{n} \left[ \int d\lambda_{k} \rho_{k}(\lambda_{k})  \sum\limits_{i=1}^{2^{m-1}} |J^{n,k}_{m,i}|\right]^{\frac{1}{n}}
	\end{equation}
	As in Eq. (\ref{alm1}), by putting $\sum\limits_{i=1}^{2^{m-1}}|J^{n,k}_{m,i}|= \alpha^{k}_{m}$ for $k\in[n]$  in  Eq. (\ref{deltamn}), and integrating over $\lambda_{k}$, we finally get
	\begin{equation}
	\label{Deltanmalpha}
	(\Delta^{n}_{m})_{nl}\leq\prod\limits_{k=1}^{n}(\alpha_m^k)^{\frac{1}{n}} =\alpha_{m} 
	\end{equation}
	Since we have used the same encoding scheme for each Alice, thus we  have $\forall k, \alpha_m^k=\alpha_m$.  Eq. (\ref{Deltanmalpha}) provides the family of generalized  $n$-locality inequalities for any arbitrary $n\geq2$ and $m\geq 2$. The value of $\alpha_m$ is given in Eq. (\ref{alpham}) and explicitly  derived in Appendix \ref{B}.
	
	To find the optimal quantum value of the expression $(\Delta_{m}^{n})$ in Eq.(\ref{Deltanmbl}),  we use an improved version of  sum-of-squares (SOS) approach,  so that, $(\Delta_{m}^{n})_{Q}\leq \beta_{m}^{n}$  for all possible quantum states $\rho_{A_{k}B}$  and measurement operators $A^{k}_{x_{k}}$ and $B_{i}$. This is equivalent to showing that there is a positive semi-definite operator $\gamma^{n}_{m}\geq 0$, which can be expressed as $\langle \gamma_{m}^{n}\rangle_{Q}=-(\Delta^{n}_{m})_{Q}+\beta^{n}_{m}$. This can be proven by considering a set of suitable positive operators $M^{n}_{m,i}$ which are polynomial functions of   $A^{k}_{x_{k}}$ and $B_{i}$, such that 
	\ba
	\label{gammanm}
	\langle\gamma^{n}_{m}\rangle=\sum\limits_{i=1}^{2^{m-1}}  \frac{{(\omega^{n}_{m,i}})^{\frac{1}{n}}}{2 }\langle\psi|(M^{n}_{m,i})^{\dagger}M^{n}_{m,i}|\psi\rangle
	\ea
	where $\omega^{n}_{m,i}$ is a positive number with $\omega^{n}_{m,i}=\prod \limits _{k=1}^{n}\omega^{n,k}_{m,i}$. The optimal quantum value of $(\Delta^{n}_{m})_{Q}$ is obtained if $\langle \gamma_{m}^{n}\rangle_{Q}=0$, i.e., 
	\begin{align}
	\label{mmnm}
	\forall i, \ \ \ M^{n}_{m,i}|\psi\rangle_{A_{1}A_{2}\cdots A_{n}B}=0
	\end{align}
where $|\psi\rangle_{A_{1}A_{2}\cdots A_{n} B}=|\psi\rangle_{A_{1}B}\otimes |\psi\rangle_{A_{2}B}\otimes . . . . \otimes |\psi\rangle_{\cdots A_{n}B}$ and $|\psi\rangle_{A_{k}B}$s are originating from independent sources $S_{k}$. By keeping in mind the expression given by Eq. (\ref{jnn}), the operators $M^{n}_{m,i}$ are suitably chosen as 
	\begin{eqnarray}
	\nonumber
	M^{n}_{m,i}|\psi\rangle&=&\prod\limits_{k=1}^{n}\frac{1}{(\omega^{n,k}_{m,i})^\frac{1}{n}}\bigg|\bigg[\sum\limits_{x_{k}=1}^{m}(-1)^{y^{i}_{x_{k}}} A_{x_{k}}^{k}\hspace{1pt}\bigg]|\psi\rangle\bigg|^{\frac{1}{n}}-|B_{i}|\psi\rangle|^{\frac{1}{n}}\\
	\label{nmi}
	&=&\prod\limits_{k=1}^{n}\frac{1}{(\omega^{n,k}_{m,i})^\frac{1}{n}}|J^{n,k}_{m,i}|\psi\rangle|^{\frac{1}{n}}-|B_{i}|\psi\rangle|^{\frac{1}{n}}
	\end{eqnarray}
	Here, for notational convenience we write $|\psi\rangle_{A_{1}A_{2}\cdots A_{n} B}=|\psi\rangle$. Note that $		\omega^{n,k}_{m,i}=||J^{n,k}_{m,i}||_2=||\sum\limits_{x_{k}=1}^{n} (-1)^{y^i_{x_{k}}} A^{k}_{x_{k}}\ket{\psi}||_{2}$ and 
	\begin{align}
	\label{Cymi}
			(\omega^{n,k}_{m,i})^{2}=\langle\psi| (J_{m,i}^{n,k})^{\dagger}(J_{m,i}^{n,k})|\psi\rangle= \langle\psi| (J_{m,i}^{n,k})^{2}|\psi\rangle.
	\end{align}
	We can then write,
\begin{eqnarray}
\label{Ckkmi}
\nonumber
&&\langle\psi|(M^{n}_{m,i})^{\dagger}M^{n}_{m,i}|\psi\rangle=\prod\limits_{k=1}^{n}\frac{1}{(\omega_{m,i}^{n,k})^\frac{2}{n}}\left(\langle\psi|\left(J_{m,i}^{n,k}\right)^{2}|\psi\rangle\right)^{\frac{1}{n}}\\
\nonumber
&&-2\bigg|\prod\limits_{k=1}^{n}\frac{1}{(\omega_{m,i}^{n,k})}\langle\psi|(J_{m,i}^{n,k})B_{i}|\psi\rangle\bigg|^{\frac{1}{n}}+\left(\langle\psi|(B_{i})^{2}|\psi\rangle\right)^{\frac{1}{n}}\\
&=&
2-2\bigg|\prod\limits_{k=1}^{n}\frac{1}{(\omega_{m,i}^{n,k})}\langle\psi|(J_{m,i}^{n,k})B_{i}|\psi\rangle\bigg|^{\frac{1}{n}}
\end{eqnarray}
where we have used Eq. (\ref{Cymi}) and $(A^{k}_{x_{k}})^{\dagger}A^{k}_{x_{k}}=B_{i}^{\dagger} B_{i}=\mathbb{I}$. Putting Eq. (\ref{Ckkmi}) in Eq. (\ref{gammanm}) we get
 \begin{eqnarray}
     \langle\gamma^{n}_{m}\rangle_{Q}&&=\sum\limits_{i=1}^{2^{m-1}}(\omega^{n}_{m,i})^{\frac{1}{n}}-\sum\limits_{i=1}^{2^{m-1}}(\omega^{n}_{m,i})^{\frac{1}{n}}\bigg|\prod\limits_{k=1}^{n}\frac{1}{(\omega_{m,i}^{n,k})}\langle\psi|(J_{m,i}^{n,k})B_{i}|\psi\rangle\bigg|^{\frac{1}{n}}\hspace{6mm}
		\end{eqnarray}
     Since $\omega^{n}_{m,i}=\prod\limits_{k=1}^{n}\omega_{m,i}^{n,k}$ we have
     \begin{eqnarray}
     \langle\gamma^{n}_{m}\rangle_{Q}
		&&=\sum\limits_{i=1}^{2^{m-1}}(\omega^{n}_{m,i})^{\frac{1}{n}}-\sum\limits_{i=1}^{2^{m-1}}\left[\bigg|\prod\limits_{k=1}^{n}\langle\psi|(J_{m,i}^{n,k})B_{i}|\psi\rangle\bigg|^{\frac{1}{n}}\right] \\
		\nonumber
     &&=\sum\limits_{i=1}^{2^{m-1}}(\omega^{n}_{m,i})^{\frac{1}{n}}-(\Delta_{m}^{n})_{Q}
		\end{eqnarray} 
		
%which in turn provides
	
\begin{figure}[]
		\centering
		{{\includegraphics[width=1.0\linewidth]{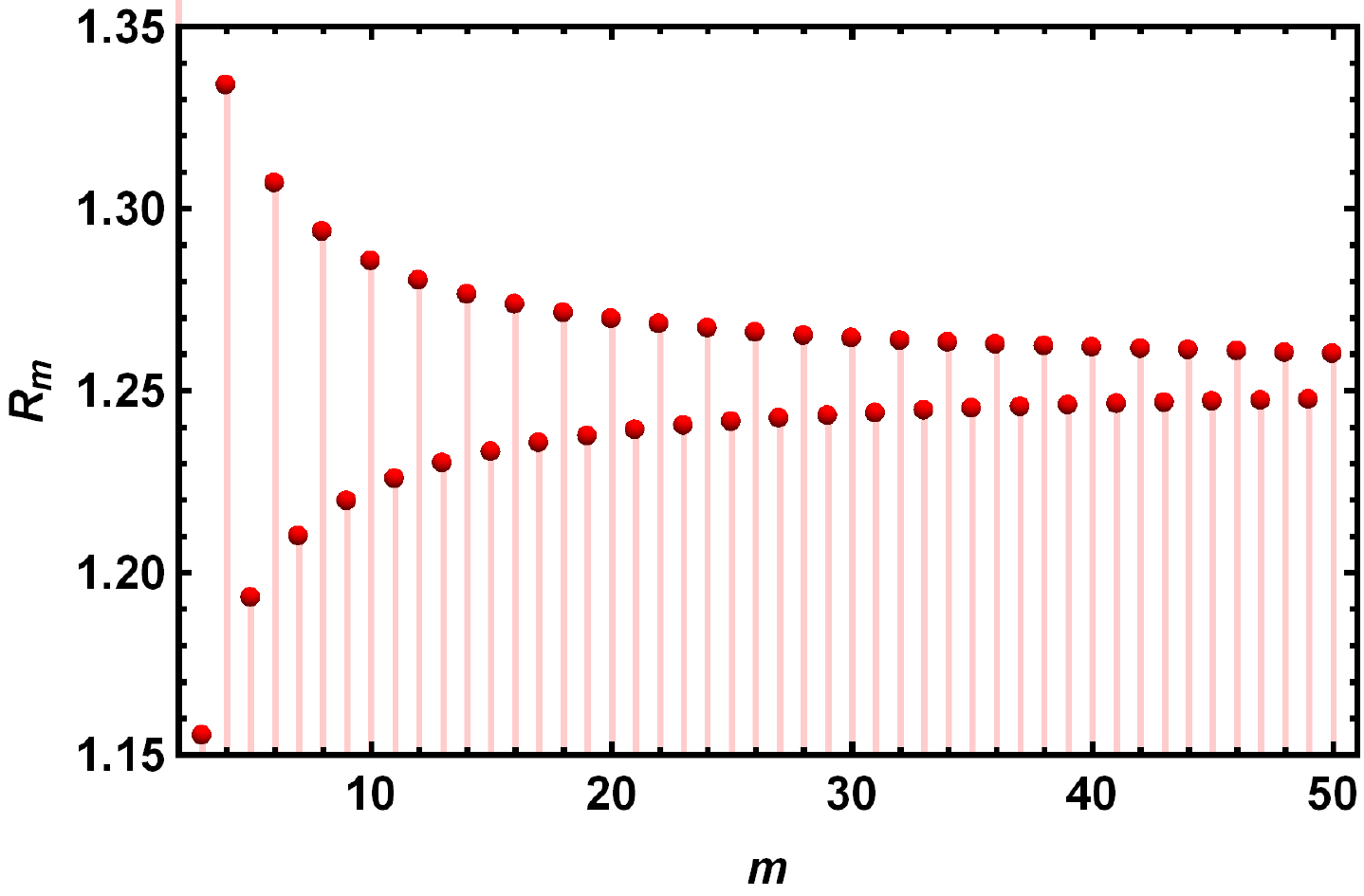}}}
		\caption{Ratio of optimal quantum value and $n$-local bound from $m=3$ to $m=50$.}
		\label{FIG. 3}
	\end{figure}

	%\begin{align}
	%\label{delopt1}
	%(\Delta^{n}_{m})_{Q}= \sum\limits_{i=1}^{2^{m-1}}(\omega^{n}_{m,i})^{\frac{1}{n}}-  \langle\gamma^{n}_{m}\rangle_{Q}
	%\end{align} 
	Since $\gamma^{n}_{m}$ is positive semi-definite, the maximum value of $	(\Delta^{n}_{m})_{Q}$ is obtained when $\langle\gamma^{n}_{m}\rangle_{Q}= 0$, i.e., 
		\begin{align}
	\label{delopt}
	(\Delta^{n}_{m})_{Q}^{opt}= max\left(\sum\limits_{i=1}^{2^{m-1}}(\omega^{n}_{m,i})^{\frac{1}{n}}\right)
	\end{align} 
	Using again  the inequality (\ref{Tavakoli}), we can write $\sum\limits_{i=1}^{2^{m-1}}(\omega^{n}_{m,i})^{\frac{1}{n}}\leq\prod \limits_{k=1}^{n}\bigg(\sum\limits_{i=1}^{2^{m-1}}\omega^{n,k}_{k,i}\bigg)^{\frac{1}{n}}$ and since $A^{k}_{x_{k}}$s are dichotomic, the quantity $\omega^{n,k}_{m,i}$ can explicitly be written as 
\begin{align}
\label{omega}
&\omega^{n,k}_{m,i}= \Big[m+\Big\langle\{(-1)^{y^{i}_{1}}A^{k}_{1},
\sum\limits_{x_{k}=2}^{m}(-1)^{y^{i}_{x_{k}}} A^{k}_{x_{k}}\}]\Big\rangle+\Big\langle\{(-1)^{y^i_2} A^{k}_{2},\\
\nonumber &\sum\limits_{x_{k}=3}^{m}(-1)^{y^{i}_{x_{k}}} A^{k}_{x_{k}}\}\Big\rangle
\cdots
+\Big\langle\{(-1)^{y^i_{m-1}} A^{k}_{(m-1)}, (-1)^{y^i_m} A^{k}_{m}\}\Big\rangle\Big]^{1/2}.
\end{align}
Following the procedure adopted for $m=3$ in Appendix A, we find $\omega^{n,k}_{m,i}\leq \sqrt{m}$ for every $i$ and $k$, and the equality holds only when every anticommutator is zero. We then have $\sum\limits_{i=1}^{2^{m-1}}(\omega^{n}_{m,i})^{\frac{1}{n}}\leq\prod \limits_{k=1}^{n}\bigg(\sum\limits_{i=1}^{2^{m-1}}\omega^{n,k}_{m,i}\bigg)^{\frac{1}{n}}= \prod \limits_{k=1}^{n}\bigg( 2^{m-1}\sqrt{m}\bigg)^{\frac{1}{n}}$ . The optimal quantum value of $(\Delta^{n}_{m})_{Q}$ is 
\begin{align}
\label{Copt}
(\Delta^{n}_{m})_{Q}^{opt}= 2^{m-1}\sqrt{m}
\end{align}
which is larger than $\alpha_{m}$ in Eq. (\ref{alpham}) for any arbitrary $m$, thereby implying the quantum violation of the family of $n$-locality inequalities given by Eq. (\ref{Deltanmalpha}). The ratio $\mathcal{R}_{m}=(\Delta^{n}_{m})_Q^{Opt}/(\Delta^{n}_{m})^{Opt}_{nl}$ between quantum and classical upper bounds for a given $m$ is plotted in Fig. \ref{FIG. 3}. We find that the ratio saturates to $1.25$ for sufficiently large values of $m$.  

As already mentioned, the optimal quantum value is obtained when for every $k$, Alice$_{k}$ chooses $m$ number of anticommuting observables ($A^{k}_{x_{k}})$. Bob's observables can be fixed from Eq. (\ref{mmnm}), so that, $B_{i}|\psi\rangle=\prod\limits_{k=1}^{n}\frac{1}{\sqrt{m}} J^{n,k}_{m,i}|\psi\rangle$. This in turn implies that the required state $|\psi\rangle$ has to satisfy $B_{i}\otimes B_{i}|\psi\rangle=|\psi\rangle$ for every $i$.

	Note here that for $m=2$ and $m=3$ cases, optimal quantum value can be achieved when every Alice shares a single two-qubit maximally entangled state with Bob. However,  at most three mutually anticommuting observables are available for qubit system. Hence, for $m\geq 4$, every Alice requires higher dimensional system to achieve the optimal value. Note that, there exists $2^{l}+1$ mutually anticommuting observables  for $l$ ($l\in \mathbb{N}$) qubit system  \cite{zei2002}. We find that to obtain the optimal value for arbitrary $m$, the local dimension of every Alice to be $d=2^{\lfloor m/2\rfloor}$. In other words, every  Alice shares at least $\lfloor m/2\rfloor$ copies of two-qubit maximally entangled state with Bob. The total state for $n+1$ parties can be written as 
\begin{equation}
\label{state}
|\psi_{A_{1}A_{2}\cdots A_{n}B}\rangle = |\phi^{+}_{A_{1}B}\rangle^{\otimes\lfloor{\frac{m}{2}\rfloor}}\otimes|\phi^{+}_{A_{2}B}\rangle^{\otimes\lfloor{\frac{m}{2}\rfloor}}\otimes\cdots\otimes|\phi^{+}_{A_{n}B}\rangle^{\otimes\lfloor{\frac{m}{2}\rfloor}} 
\end{equation}
 where $|\phi^{+}_{A_{k}B}\rangle^{\otimes\lfloor{\frac{m}{2}}}$ is the state originated from each independent source $S_{k}$. The above discussion clearly indicates the possibility of witnessing the dimension of Hilbert space in network scenario and activation of non $n$-locality by using multiple copies of entangled states.

\section{Summary and discussion} 
\label{V}
 In summary, we have provided a non-trivial generalization of the $n$-locality scenario in the star-network configuration. Such a network involves $n$ number of edge observers (Alices), a central observer (Bob), and $n$ number of independent sources. Instead of two dichotomic measurements per party, we considered that every Alice performs an arbitrary $m$ number of dichotomic measurements, and the central observer Bob performs $2^{m-1}$ number of dichotomic measurements. We  derived a family of generalized $n$-locality inequalities.

As a case study, we first considered the bilocality scenario for $m=3$ case and derived the bilocality inequality, when each of the two Alices performs three binary-outcome measurements, and Bob performs four binary-outcome measurements.  We demonstrated the optimal quantum violation using an improved version of the usual sum-of-squares (SOS) approach. The optimal quantum value is attained when observables of each Alice are mutually anticommuting. Bob's observables and the required entangled states are then fixed by the optimization condition. We showed that each Alice has to share a two-qubit maximally entangled state with Bob to obtain the optimal value.  We further extended the argument to $n$-locality scenario in star-network configuration.

We then extended our study for arbitrary $ m$ and derived a family of generalized $n$-locality inequalities in star-network configuration. We demonstrated that the optimal quantum violation is attained when all $m$ observables of every Alice has to be mutually anticommuting. Since there are at most three mutually anticommuting observables for a qubit system, for $m>3$, the local dimension of Hilbert space of each Alice needs has to be more than two. We found that the local dimension of every Alice needs to be $d=2^{\lfloor m/2\rfloor}$. We argued that every Alice and Bob share $\lfloor m/2\rfloor$ copies of two-qubit maximally entangled state to achieve the optimal quantum violation of the family of generalized $n$-locality inequalities. 

Hence, a single copy of a two-qubit entangled state may not reveal the non-$n$-locality for $m> 3$ cases, but multiple copies of it may activate non-$n$-locality  in the network. Such a feature indicates witnessing the dimension of Hilbert space in network scenario. Suppose for $m=4$,  we find $(\Delta^{n}_{4})_{Q} > (\Delta^{n}_{4})_{nl}$ for a single two-qubit maximally entangled state, thereby providing an upper bound for qubit system. To obtain the optimal value $(\Delta^{n}_{4})_{Q}^{opt}$, a pair of two-qubit maximally entangled state is required to be shared by $k^{th}$ Alice and Bob,  where $k\in [n]$. Instead of a two-qubit maximally entangled state, one may use a noisy two-qubit entangled state of the form  $\rho_{A_{k}B}(v_{k})=v_{k}|\psi_{k}\rangle\langle\psi_{k}|+(1-v_{k}) \mathbb I/4$ where  $v_{k}$ is the visibility with $k\in [n]$ . In such a case it may be possible that $(\Delta^{n}_{4})_{Q} < (\Delta^{n}_{4})_{nl}$ when the value of visibility parameter $v_{k}$ is lower than the critical value $v_{k}^{\ast}$. Now, if every Alice and Bob share a pair of noisy two-qubit entangled states, then the non-$n$-locality in network  may be activated. This argument holds for any arbitrary $m>4$. 
Detailed study of witnessing the Hilbert space dimension and activation of non-$n$-locality across the network for multiple copies of entangled states could be an interesting open problem to study.  As an offshoot of our work, studying various other topologies of quantum networks for arbitrary $m$ could be another interesting line of work.

\emph{Aknowledgement:-} SM acknowledges the research grant
SB/S2/RJN-083/2014.  RK acknowledges the UGC fellowship (F.No. 16-6(Dec.2017)/2018(NET/CSIR)). AKP acknowledges the support from
the project DST/ICPS/QuST/Theme-1/2019/4.

   	\begin{widetext}
		\appendix
		\section{Optimal value of bilocality inequality $(n=2)$ for $m=3$}
		\label{A}

To derive the optimal quantum value of $(\Delta^{2}_{3})_{Q}$, we use a variant of the known sum-of-squares (SOS) approach. We show that there is a positive semidefinite operator $\gamma^{2}_{3}$, that can be expressed as $\langle \gamma^2_{3}\rangle_{Q}=-(\Delta^{2}_{3})_{Q}+\beta^{2}_{3}$. Here $\beta^{2}_{3}$ is the optimal value, can be obtained when $\langle \gamma^2_{3}\rangle_{Q}$ is equal to zero. This can be proven by considering a set of suitable positive operators $M^{2}_{3,i}$ which is polynomial functions of   $A^{1}_{x_{1}}$, $B_{i}$ and $A^{2}_{x_{2}}$ so that 
	\begin{equation}
	\label{Agamma3}
	\langle\gamma^{2}_{3}\rangle=\sum\limits_{i=1}^4\dfrac{\sqrt{\omega_{3,i}^{2}}}{2}\langle\psi|(M^{2}_{3,i})^{\dagger}(M^{2}_{3,i})|\psi\rangle\end{equation}
	where $\omega^{2}_{3,i}$ is suitable positive numbers and $\omega^{2}_{3,i}=\omega^{2,1}_{3,i}\cdot\omega^{2,2}_{3,i}$.  	We choose a  suitable set of  positive operators $M^{2}_{3,i}$ as
\ba
\label{a2}
    M^{2}_{3,1}|\psi\rangle_{A_{1}BA_{2}}=\sqrt{\bigg|\left(\frac{{A}^{1}_{1}+{A}^{1}_{2}+A^{1}_{3}}{(\omega_{3,1}^{2,1})}\otimes\frac{{A}^{2}_{1}+{A}^{2}_{2}+A^{2}_{3}}{(\omega_{3,1}^{2,2})}\right) |\psi\rangle_{A_{1}BA_{2}}
	\bigg|} -\sqrt{|B_{1}|\psi\rangle_{A_{1}BA_{2}}|}\\
		M^{2}_{3,2}|\psi\rangle_{A_{1}BA_{2}}=\sqrt{\bigg|\left(\frac{{A}^{1}_{1}+{A}^{1}_{2}-A^{1}_{3}}{(\omega_{3,2}^{2,1})}\otimes\frac{{A}^{2}_{1}+{A}^{2}_{2}-A^{2}_{3}}{(\omega_{3,2}^{2,2})}\right) |\psi\rangle_{A_{1}BA_{2}}	\bigg|} -\sqrt{|B_{2}|\psi\rangle_{A_{1}BA_{2}}|}\\
	M^{2}_{3,3}|\psi\rangle_{A_{1}BA_{2}}=\sqrt{\bigg|\left(\frac{{A}^{1}_{1}-{A}^{1}_{2}+A^{1}_{3}}{(\omega_{3,3}^{2,1})}\otimes\frac{{A}^{2}_{1}-{A}^{2}_{2}+A^{2}_{3}}{(\omega_{3,3}^{2,2})}\right) |\psi\rangle_{A_{1}BA_{2}}
	\bigg|} -\sqrt{|B_{3}|\psi\rangle_{A_{1}BA_{2}}|}\\
	\label{a4}
		M^{2}_{3,4}|\psi\rangle_{A_{1}BA_{2}}=\sqrt{\bigg|\left(\frac{-{A}^{1}_{1}+{A}^{1}_{2}+A^{1}_{3}}{(\omega_{3,4}^{2,1})}\otimes\frac{-{A}^{2}_{1}+{A}^{2}_{2}+A^{2}_{3}}{(\omega_{3,4}^{2,2})}\right) |\psi\rangle_{A_{1}BA_{2}}
	\bigg|} -\sqrt{|B_{4}|\psi\rangle_{A_{1}BA_{2}}|}
	\ea	
	where $\omega^{2,k}_{3,1}$ is defined as $\omega^{2,k}_{3,1}=||(A^{k}_{1}+A^{k}_{2}+A^{k}_{3})|\psi\rangle_{A_1BA_2}||_{2}$ and similarly for other  $\omega^{2,k}_{3,i}$ s, $k=1,2$ . Here $||.||_{2}$ denotes the norm of the vector. 
	Plugging Eqs. (\ref{a2}-\ref{a4}) in Eq. (\ref{Agamma3}), we get 
	$\langle\gamma^{2}_{3}\rangle_{Q}=-(\Delta^{2}_{3})_{Q} +\sum\limits_{i=1}^{4}\sqrt{\omega^{2}_{3,i}}$. 	The optimal quantum value of $(\Delta^{2}_{3})_{Q}$ is obtained if $\langle \gamma^{2}_{3}\rangle_{Q}=0$, implying that	$ \forall i, M^{2}_{3,i}|\psi\rangle_{A_{1}BA_{2}}=0$.	Hence,
	\begin{eqnarray}(\Delta^{2}_{3})_{Q}^{opt} =max\left(\sum\limits_{i=1}^{4}\sqrt{\omega^{2}_{3,i}}\right)
	\end{eqnarray}
	As defined, 	$\omega^{2,1}_{3,1}=||(A^{1}_{1}+A^{1}_{2}+A^{1}_{3})|\psi\rangle_{A_1BA_2}||_{2}=\sqrt{3+\langle(\{A^{1}_{1},A^{1}_{2}\}+\{A^{1}_{2},A^{1}_{3}\}+\{A^{1}_{3},A^{1}_{1}\})\rangle}.$ 	Similarly,  we can write
	\ba
	\nonumber
	&&\omega^{2,2}_{3,1}=\sqrt{3+\langle(\{A^{2}_{1},A^{2}_{2}\}+\{A^{2}_{2},A^{2}_{3}\}+\{A^{2}_{3},A^{2}_{1}\})\rangle},\hspace{5pt}
	\omega^{2,1}_{3,2}=\sqrt{3+\langle(\{A^{1}_{1},A^{1}_{2}\}-\{A^{1}_{2},A^{1}_{3}\}-\{A^{1}_{3},A^{1}_{1}\})\rangle}.\\
		&&\omega^{2,2}_{3,2}=\sqrt{3+\langle(\{A^{2}_{1},A^{2}_{2}\}-\{A^{2}_{2},A^{2}_{3}\}-\{A^{2}_{3},A^{2}_{1}\})\rangle}\hspace{5pt},
\omega^{2,1}_{3,3}=\sqrt{3+\langle(-\{A^{1}_{1},A^{1}_{2}\}-\{A^{1}_{2},A^{1}_{3}\}+\{A^{1}_{3},A^{1}_{1}\})\rangle}.\\
	\nonumber
		&&\omega^{2,2}_{3,3}=\sqrt{3+\langle(-\{A^{2}_{1},A^{2}_{2}\}-\{A^{2}_{2},A^{2}_{3}\}+\{A^{2}_{3},A^{2}_{1}\})\rangle},\hspace{5pt} \omega^{2,1}_{3,4}=\sqrt{3+\langle(-\{A^{1}_{1},A^{1}_{2}\}+\{A^{1}_{2},A^{1}_{3}\}-\{A^{1}_{3},A^{1}_{1}\})\rangle}.\\
		\nonumber
	&&\omega^{2,2}_{3,4}=\sqrt{3+\langle(-\{A^{2}_{1},A^{2}_{2}\}+\{A^{2}_{2},A^{2}_{3}\}-\{A^{2}_{3},A^{2}_{1}\})\rangle}.
	\ea
	
	Since  $\omega^{2}_{3,i}=\omega^{2,1}_{3,i}\cdot\omega^{2,2}_{3,i}$, by using the  inequality $\sum\limits_{i=1}^{4}\sqrt{r_{i}s_{i}}\leq\sqrt{\sum\limits_{i=1}^{4} r_{i}}\sqrt{\sum\limits_{i=1}^{4}s_{i}}$  (for $r_{i}, s_{i} \geq 0$,  $i=1,2,3,4)$, we get $\sum\limits_{i=1}^{4}\sqrt{\omega^{2}_{3,i}}\leq \sqrt{\sum\limits_{i=1}^{4}\omega^{2,1}_{3,i}}\sqrt{\sum\limits_{i=1}^{4}\omega^{2,2}_{3,i}}$  .   By using the identity 	$\sqrt{b+a}+\sqrt{b-a}=\sqrt{2b+2\sqrt{b^{2}-a^{2}}}$, we obtain
		\begin{eqnarray}
		\nonumber
		\sum\limits_{i=1}^{4}\sqrt{\omega^{2}_{3,i}}&\leq& \Bigg(\sqrt{2\bigg(3+\langle\{A^{1}_{1},A^{1}_{2}\}\rangle\bigg)+2\sqrt{\bigg(3+\langle\{A^{1}_{1},A^{1}_{2}\}\rangle\bigg)^{2}-\bigg(\langle\{A^{1}_{2},A^{1}_{3}\}\rangle+\langle\{A^{1}_{3},A^{1}_{1}\}\rangle\bigg)^{2}}}\\
		\nonumber
		&+&\sqrt{2\bigg(3-\langle\{A^{1}_{1},A^{1}_{2}\}\rangle\bigg)+2\sqrt{\bigg(3-\langle\{A^{1}_{1},A^{1}_{2}\}\rangle\bigg)^{2}-\bigg(\langle\{A^{1}_{2},A^{1}_{3}\}\rangle-\langle\{A^{1}_{3},A^{1}_{1}\}\rangle\bigg)^{2}}}\hspace{5pt}\Bigg)^{\frac{1}{2}}\\
		&&\cdot \Bigg(\sqrt{2\bigg(3+\langle\{A^{2}_{1},A^{2}_{2}\}\rangle\bigg)+2\sqrt{\bigg(3+\langle\{A^{2}_{1},A^{2}_{2}\}\rangle\bigg)^{2}-\bigg(\langle\{A^{2}_{2},A^{2}_{3}\}\rangle+\langle\{A^{2}_{3},A^{2}_{1}\}\rangle\bigg)^{2}}}\\
		\nonumber
		&+&\sqrt{2\bigg(3-\langle\{A^{2}_{1},A^{2}_{2}\}\rangle\bigg)+2\sqrt{\bigg(3-\langle\{A^{2}_{1},A^{2}_{2}\}\rangle\bigg)^{2}-\bigg(\langle\{A^{2}_{2},A^{2}_{3}\}\rangle-\langle\{A^{2}_{3},A^{2}_{1}\}\rangle\bigg)^{2}}}\hspace{5pt}\Bigg)^{\frac{1}{2}}
		\end{eqnarray}
	Clearly,  $(\Delta^2_3)^{opt}_Q=max\left({\sum\limits_{i=1}^{4}\sqrt{\omega^{2}_{3,i}}}\right)=4\sqrt{3}$. This  is obtained when  $A^{1}_{1},A^{1}_{2}$ and $A^{1}_{3}$ are mutually anticommuting and same for $A^{2}_{1},A^{2}_{2}$ and $A^{2}_{3}$. It is straightforward to find Bob's observables from $M^2_{3,i}|\psi\rangle_{A_1BA_2}=0$, $\forall i$ which, in turn, fixes the state $|\psi\rangle_{A_1BA_2}$ to be a two-qubit maximally entangled state.	The same approach is used in the main text to derive the optimal quantum violations of the family of $n$-local inequalities for arbitrary $m$.
		\section {Derivation of $\alpha_{m}$ in Eq. (\ref{Deltanmalpha}) of the main text}
		\label{B}
	The $n$-locality upper bound $\alpha_{m}^{k}$ for a given $k$ is given by

 \begin{eqnarray}
	\label{alm}
	\alpha_{m}^{k}= \sum_{i=1}^{2^{m-1}}\left|J_{m,i}^{n,k}\right|=\sum_{i=1}^{2^{m-1}}\bigg|\sum\limits_{x_{k}=1}^{m}(-1)^{y_{x_{k}}^{i}}\langle{A^{k}_{x_{k}}}\rangle_{\lambda_{n}}\bigg|
	\end{eqnarray}
	As mentioned in the main text, for a given $i\in \{1,2,\cdots ,2^{m-1}\}$, the quantity ${y_{x_{k}}^{i}}$ is either $0$ or $1$, fixed by the encoding scheme of random-access code. Here $y^{i}$ contains those elements (bit strings) of $y^{\delta}\in \{0,1\}^{m}$ having first bit $0$.  The term ${y_{x_{k}}^{i}}$ then denotes the $x_{k}^{th}$ bit of $y^{i}$. By writing the length $m$ bit string $y^{i}$ as $i^{th}$ column we have the generator matrix of the augmented Hadamard code \cite{had} of order $m\times 2^{m-1}$ as 
\begin{center}
	
	$G=\begin{bmatrix}\bigg\uparrow&	\bigg\uparrow&\bigg\uparrow&\cdots &\bigg\uparrow\\
	
	y^1&	y^2&y^3&\cdots &y^{2^{m-1}}\\
	\bigg\downarrow&\bigg\downarrow&\bigg\downarrow&\cdots &\bigg\downarrow
	\end{bmatrix}=
	\begin{bmatrix}
	y^1_{x_1=1}&	y^2_{x_1=1}&y^3_{x1=1}&\cdots &y^{2^{m-1}}_{x_1=1}\\
	\\
	y^1_{x_1=2}&	y^2_{x_1=2}&y^3_{x_1=2}&\cdots &y^{2^{m-1}}_{x_1=2}\\
	
	\vdots&\vdots&\vdots&\ddots&\vdots\\
	
		y^1_{x_1=m-1}&	y^2_{x_1=m-1}&y^3_{x_1=m-1}&\cdots &y^{2^{m-1}}_{x_1=m-1}\\
		\\
		y^1_{x_1=m}&	y^2_{x_1=m}&y^3_{x_1=m}&\cdots &y^{2^{m-1}}_{x_1=m}
	\end{bmatrix}_{m\times 2^{m-1}}=	\begin{bmatrix}0&	0&0&\cdots &0\\
	\\
	0&	0&0&\cdots &1\\
	\\
	\vdots&\vdots&\vdots&\ddots&\vdots\\
	0&0&1&\cdots &1\\
	\\
	0&1&0&\cdots &1
	\end{bmatrix}_{m\times 2^{m-1}}$
\end{center}
	
	Since $J_{m,i}^{n,k}=\sum\limits_{x_{k}=1}^{m}(-1)^{y_{x_{k}}^{i}}\langle{A^{k}_{x_{k}}}\rangle_{\lambda_{k}}$, then there is a one-to-one correspondence with the $i^{th}$ column of $G$. We can then write

	\begin{center}
	$J^{k}=	\begin{bmatrix}\bigg\uparrow&	\bigg\uparrow&\bigg\uparrow&\cdots &\bigg\uparrow\\
	\\
	
	J_{m,1}^{k}&	J_{m,2}^{k}&J_{m,3}^{k}&\cdots &J_{m,2^{m-1}}^{k}\\\\
	\bigg\downarrow&\bigg\downarrow&\bigg\downarrow&\cdots &\bigg\downarrow
	\end{bmatrix}=\begin{bmatrix}
		\left\langle A_{1}^{k} \right\rangle_{\lambda_{k}} &	\left\langle A_{1}^{k} \right\rangle_{\lambda_{k}} & \left\langle A_{1}^{k} \right\rangle_{\lambda_{k}}&\cdots & \left\langle A_{1}^{k}\right\rangle_{\lambda_{k}} \\
		\\
	\left\langle A_{2}^{k}\right\rangle_{\lambda_{k}}&	\left\langle A_{2}^{k}\right\rangle_{\lambda_{k}}& \left\langle A_{2}^{k}\right\rangle_{\lambda_{k}}&\cdots &-\left\langle A_{2}^{k}\right\rangle_{\lambda_{k}}\\
		\vdots&\vdots&\vdots&\ddots&\vdots\\
		\left\langle A_{m}^{k}\right\rangle_{\lambda_{k}}&	-\left\langle A_{m}^{k}\right\rangle_{\lambda_{k}}& \left\langle A_{m}^{k}\right\rangle_{\lambda_{k}}&\cdots &-\left\langle A_{m}^{k}\right\rangle_{\lambda_{k}}
	\end{bmatrix}_{m\times 2^{m-1}}$
\end{center}

 For deriving $\alpha_{m}^{k}$, we require the modulus of each column of $J^{k}$ and their sum.   We also note that $\left\langle A_{x_k}^{k} \right\rangle_{\lambda_{k}}=\pm 1$. A bit-flip operation in a row of $G$ corresponds to the change of sign of $\left\langle A_{x_k}^{k} \right\rangle_{\lambda_{k}}$ in the same row of $J^{k}$. There are $2^{m}$ number of permutations of bit-flips are possible for a given $m$. However, the encoding scheme in a random-access-code is so constructed that due to sign change of observables, the modulus value of each column $|J_{m,i}^{n,k}|$ may change its place but for every permutation of observable signs just corresponds to the permutation of column.  Hence, for every permutation of observable signs,  $\sum_{i=1}^{2^{m-1}}\left|J_{m,i}^{n,k}\right|$ remains invariant. 

Hence, without loss of generality, we can derive the upper bound $\alpha_{m}^{k}$ by simply taking the outcomes of all observables are $+1$. Now, note that there are total $m\cdot2^{m-1}$ number of $\pm 1$ entries in $2^{m-1}$ number of columns in $J^k$. Among them, $\binom{m}{1}$ columns have one $-1$ entry , $\binom{m}{2}$ columns have two $-1$ entries, $\cdots$ $\binom{m}{\left\lfloor\frac{m}{2}\right \rfloor-1}$ columns have $\bigg(\left\lfloor\frac{m}{2}\right \rfloor-1\bigg)$ number of $-1$ entries. Now, if $m$ is odd, then $\binom{m}{\left\lfloor\frac{m}{2}\right\rfloor}$ columns have $\left\lfloor\frac{m}{2}\right\rfloor$ number of $-1$ entries. But, if $m$ is even, then  $\frac{1}{2}\binom{m}{\left\lfloor\frac{m}{2}\right \rfloor}$ columns have $\left\lfloor\frac{m}{2}\right \rfloor$ number of  $-1$ entries.  Altogether, by taking sum of the modulus of the sum of each column, we have   \begin{equation}
 \label{Balm}
     \alpha^{k}_{m}=m 2^{m-1}-2\left(\sum\limits_{j=1}^{\left\lfloor{\frac{m}{2}}\right \rfloor-1} j\binom{m}{j} + \left(1- \frac{(m \oplus 1)}{2}\right)\left\lfloor{\frac{m}{2}}\right \rfloor\binom{m}{\left\lfloor{\frac{m}{2}}\right\rfloor }\right)\end{equation}\\
 Now, in order to match the form in last quantity, we  re-write $m2^{m-1}$ in a specific form as 
\begin{eqnarray}
\label{b3}
 m2^{m-1}&=&\sum\limits_{j=0}^{\lfloor\frac{m}{2}\rfloor-1}m\binom{m}{j}+\left(1- \frac{(m \oplus 1)}{2}\right)m\binom{m}{\left\lfloor{\frac{m}{2}}\right \rfloor}
 \end{eqnarray}
 Using Eq. (\ref{b3}) in Eq.(\ref{Balm}), we get the following. If $m$ is odd, we have
 \begin{eqnarray}
\alpha_{m}^{k}=\sum\limits_{j=0}^{\lfloor\frac{m}{2}\rfloor-1}m\binom{m}{j}+m\binom{m}{\left\lfloor{\frac{m}{2}}\right \rfloor}-2\left(\sum\limits_{j=1}^{\left\lfloor{\frac{m}{2}}\right \rfloor-1} j\binom{m}{j} + \left\lfloor{\frac{m}{2}}\right \rfloor\binom{m}{\left\lfloor{\frac{m}{2}}\right\rfloor }\right)=\sum\limits_{j=0}^{\lfloor\frac{m}{2}\rfloor-1}\binom{m}{j}(m-2j)+\binom{m}{\left\lfloor{\frac{m}{2}}\right \rfloor}\left(m-2\left\lfloor\frac{m}{2}\right\rfloor\right)\equiv\sum\limits_{j=0}^{\lfloor\frac{m}{2}\rfloor}\binom{m}{j}(m-2j)\end{eqnarray} and if $m$ is even, we have

\begin{eqnarray} 
\alpha_{m}^{k}=\sum\limits_{j=0}^{\lfloor\frac{m}{2}\rfloor-1}m\binom{m}{j}+\dfrac{m}{2}\binom{m}{\left\lfloor{\frac{m}{2}}\right \rfloor}-2\left(\sum\limits_{j=1}^{\left\lfloor{\frac{m}{2}}\right \rfloor-1} j\binom{m}{j} + \frac{1}{2}\left\lfloor{\frac{m}{2}}\right \rfloor\binom{m}{\left\lfloor{\frac{m}{2}}\right\rfloor }\right)=\sum\limits_{j=0}^{\lfloor\frac{m}{2}\rfloor-1}\binom{m}{j}(m-2j)+\binom{m}{\left\lfloor{\frac{m}{2}}\right \rfloor}\left(m-2\left\lfloor\frac{m}{2}\right\rfloor\right)\equiv\sum\limits_{j=0}^{\lfloor\frac{m}{2}\rfloor}\binom{m}{j}(m-2j)
\end{eqnarray}

Thus, for any arbitrary $m$, 
\begin{eqnarray}
\alpha_{m}^{k}&=&\sum\limits_{j=0}^{\lfloor\frac{m}{2}\rfloor}\binom{m}{j}(m-2j)
\end{eqnarray}
As we argued in the main text, $\alpha^{k}_{m}$ is same for every $k$,  we then have $\Delta_{m}^{n}\leq \prod\limits_{k=1}^{n}\bigg(\alpha_{m}^{k}\bigg)^{\frac{1}{n}}=\alpha_{m}$. This is placed in Eq.(\ref{alpham}) in the main text.

		\end{widetext}

\appendix

\end{document}